\documentclass[proof]{pasj00}
\draft

\begin{document}
\SetRunningHead{Kawashima et al}{New Supercritical State}
\Received{2000/12/31}
\Accepted{2001/01/01}

\title{New Spectral State of Supercritical Accretion Flow with Comptonizing Outflow}

\author{Tomohisa \textsc{Kawashima},\altaffilmark{1,2}%
Ken \textsc{Ohsuga},\altaffilmark{3,4}
Shin \textsc{Mineshige},\altaffilmark{5}
Dominikus \textsc{Heinzeller},\altaffilmark{5}
Hideaki \textsc{Takabe},\altaffilmark{6}
and
Ryoji {\sc Matsumoto}\altaffilmark{1}
}

\altaffiltext{1}{Department of Physics, Graduate School of Science,
Chiba University, 1-33 Yayoi-cho, Inage-ku, Chiba 263-8522, Japan}
\email{kawashima-t@astro.s.chiba-u.ac.jp}
\altaffiltext{2}{Department of Space and Earth, Graduate School of Osaka
University, Toyonaka, Japan}
\altaffiltext{3}{National Astronomical Observatory of Japan, 2--21--1
Osawa, Mitaka-shi, Tokyo 181--8588}
\altaffiltext{4}{Institute of Physical and Chemical Reserch (RIKEN), 2-1
Hirosawa, Wako, Saitama 351-0198, Japan}
\altaffiltext{5}{Department of Astronomy, Kyoto
University,Kitashirakawa, Sakyo-ku, Kyoto 606-8502, Japan}
\altaffiltext{6}{Institute of Laser Engineering, Osaka University, 2-6
Yamada-Oka, Suita, Osaka 565-0871, Japan}


\maketitle

\begin{abstract}
Supercritical accretion flows inevitably produce radiation-pressure driven outflows, 
which will Compton up-scatter soft photons from the underlying accretion flow, 
thereby making hard emission.  
We perform two dimensional radiation hydrodynamic simulations of supercritical accretion flows and outflows, incorporating such Compton scattering effects, 
and demonstrate that there appears a new hard spectral state 
at higher photon luminosities than that of the slim-disk state.
In this state, as the photon luminosity increases, 
the photon index decreases and the fraction of the hard emission increases.
The Compton $y$-parameter is of the order of unity 
(and thus the photon index will be $\sim 2$) 
when the apparent photon luminosity is ${\sim}30L_{\rm E}$ 
(with $L_{\rm E}$ being the Eddington luminosity) for nearly face-on sources.  
This explains the observed spectral hardening of the ULX NGC1313 X-2 
in its brightening phase and thus supports the model of 
supercritical accretion onto stellar mass black holes in this ULX.

\end{abstract}

\section{Introduction} \label {Int}
It is long known that 
astrophysical black holes (especially, black hole binaries)
exhibit several distinct spectral states.
These are, in the order of increasing photon luminosities (see, e.g., \cite {Esi97}):
a low/hard state with a hard, power-law-like emission component, 
a high/soft state with a soft, thermal emission component (e.g.,\cite{Tan72,Dol79}), 
and a very high state showing both spectral components (e.g.,\cite
{Miya91,KD04}).
In addition, there seems to exist another soft spectral state 
at even higher photon luminosities, comparable to the Eddington luminosity ($L_{\rm E}$), 
called the slim disk state (\cite {Ab88,Wa00},
see \cite {KFM08} for a review and references therein).
Although recent observations suggest its existence \citep {Vi06, Oka06},
 a robust observational proof of this state has not yet been obtained.

In this respect it is interesting to note that 
Ultraluminous X-ray sources (ULXs), which are recently found successively 
in the off-center region of nearby external galaxies, also show 
both the soft thermal and the hard power-law spectral states \citep {Mak00,Cro04}.  Some of them, such as IC342 X-1 and X-2, even show spectral transitions 
between these two states \citep {KM01}. 
Notably, the typical photon luminosities of ULXs range between 
$L_{\rm ph} \simeq 10^{39-41}$ erg s$^{-1}$, 
which exceeds the Eddington luminosity for neutron stars and
stellar-mass black holes.
There are two possible models considered to account for such large
photon luminosities: 
subcritical accretion (i.e., accretion below the Eddington accretion rate)
 onto an intermediate-mass black hole (IMBH, 
 \cite {Mil03,Mil04,Cro04,Ber08})
and supercritical accretion onto a stellar-mass black hole 
\citep {King01, Ebi03, Oka06, Tsu06, Vi06, Pou07, Vi08}.
Since the black hole masses of ULXs are poorly known,
we cannot discriminate between these two models at present.
The study of spectral states and spectral transitions, in analogy to
black hole binaries, are one possibility to resolve this issue,
but the situation turns out to be very complex.
If the two states of ULXs correspond to the low/hard state and the high/soft state, 
the photon luminosities of ULXs should be sub-Eddington and 
thus the IMBH hypothesis will be favored.
If, on the other hand, the soft state of ULXs corresponds to the
 slim-disk state, 
these ULXs should be supercritical sources, which means
that the black holes cannot be IMBHs.  
The proper identification of the spectral states is thus essential 
for understanding the nature of accretion flows and black holes in ULXs. 

Interesting trends have been reported recently.  
Some ULXs (such as HoIX X-1) show spectral softening 
when the photon luminosity increases \citep {KP08}.
It is thus natural to conclude that
the brighter, soft state corresponds to the slim-disk state.
Then, its photon luminosity should be close to the Eddington luminosity, 
thereby supporting the supercritical hypothesis for ULXs.  
To our surprise, however,
there exists another class of ULXs (such as NGC1313 X-2) which exhibit the 
completely opposite behavior; that is, 
they show spectral hardening as the photon luminosity increase 
 \citep {Rob06,Miz07}.
It is then reasonable to conclude that
the observed soft-to-hard transitions correspond to the transition
from the high/soft state to the very high state in the black hole binaries.
In this case, the photon luminosities should be sub-Eddington and, hence,
the IMBH model will be favored. 
However, we can propose another interpretation.
If a hard spectral state exists in the supercritical regime and if these
sources are in such a very bright phase, the supercritical model will
survive.
From a theoretical point of view, this seems to be a natural
consequence: supercritical accretion flows inevitably produce
radiation-pressure driven outflows and such outflows will Compton
up-scatter soft photons, thereby making a hard emission component.  
The higher the photon luminosity is, the harder emission we expect.
In this paper we will demonstrate that this is indeed feasible 
based on new two-dimensional radiation-hydrodynamic (RHD) simulations, 
which incorporate the Compton scattering effects.

Two-dimensional RHD simulations of supercritical accretion flows
 around stellar-mass black holes were pioneered by \citet {Egg88} and 
followed by several authors \citep {Oku02,Oku05,Ohs07}.
Supercritical accretion in a quasi-steady state was first
calculated by \citet {Ohs05}, who
found that the total photon luminosities can indeed exceed three times 
the Eddington luminosity, 
whereas the apparent photon luminosities become more than ten times larger 
than the Eddington luminosity for a face-on observer (see also \cite {HMO06}).
They have also shown that hot outflows with their gas temperature higher than
 $10^{9}\,\textup{K}$ appear above the disk.
This suggests that the inverse Compton scattering of soft disk emission by 
such high temperature plasmas should be important, 
though it was not yet studied quantitatively.
This motivated us to perform extended simulations
that incorporate Compton scattering effects.  
We find that
the higher the mass accretion rate is, 
more high-temperature gas exists as outflow and, therefore, 
the larger becomes the Compton $y$-parameter (see equation (\ref{yp}) in
section \ref{RES} for the definition of the Compton $y$-parameter).  
We thus expect that SED becomes harder when the photon luminosity is higher.
This new state will explain the SED variation of some ULXs.

The plan of this paper is as follows:
In $\S$\ref {ME} we present basic equations and numerical methods.
In $\S$\ref {RES} we present the structure of supercritical accretion flows
with Comptonizing outflows and compute the Compton $y$-parameter.
Finally, $\S$\ref {DIS} is devoted to discussion of our findings.


\section{Methods}\label{ME}
Following \citet {Ohs05}, but  also considering the effects of Compton scattering 
in the energy exchange between photons and electrons, 
we solve the RHD equations   
in spherical coordinates $(r,{\theta},{\phi})$.
The radiative transfer equation is solved using
 the flux-limited diffusion approximation \citep {LP81,TS01}.
The general relativistic effects are incorporated by a pseudo-Newtonian
potential,
 ${\Psi}=-GM/(r-r_{\rm s})$ \citep{Pacz80}, where $r_{\rm s}(=2GM/c^{2})$ is the Schwarzschild radius, 
$G$ is the gravitational constant,
 $M$ is the mass of the black hole ($M=10{\MO}$ is employed),
 and $c$ is the speed of light.
We assume that the flow is non-self gravitating, axisymmetric with respect to the rotation axis (i.e., ${\partial}/{\partial}{\phi}=0$), and symmetric relative to the equatorial plane (where $\theta = \pi/2$).
We also adopt the ${\alpha}$ viscosity prescription \citep {SS73} and set
 ${\alpha}~=~0.1$.

The basic equations are the same as those in \citet {Ohs05} except that we add
 Compton heating/cooling terms in the energy equations of gas and radiation;
\begin{eqnarray}
\frac{{\partial}e}{{\partial}t}+{\nabla}{\cdot}(e{\mbox{\boldmath$v$}})
&=&-p{\nabla}{\cdot}{\mbox{\boldmath$v$}}-4{\pi}{\kappa}B+c{\kappa}E_{0}
+{\Phi}_{\rm vis}-{\Gamma}_{\mathrm{Comp}}~, \\
\frac{{\partial}E_{0}}{{\partial}t}+{\nabla}{\cdot}(E_{0}{\mbox{\boldmath$v$}})
&=&-{\nabla}{\cdot}{\mbox{\boldmath$F$}_{0}}
-{\nabla}{\mbox{\boldmath$v$}}{\colon}{\mathbf{P}}_{0}
+4{\pi}{\kappa}B-c{\kappa}E_{0}+{\Gamma}_{\mathrm{Comp}}.
\end{eqnarray}
Here $\mbox{\boldmath$v$}=(v_{r},~v_{\theta},~v_{\phi})$ is the velocity,
 $p$ is the gas pressure,
 $e$ is the internal energy density of the gas,
 $B$ is the blackbody intensity,
 $E_{0}$ is the radiation energy density, where the suffix $0$
 represents quantities in the comoving frame.
 $\mbox{\boldmath$F$}_{0}=(F_{0}^{r},~F_{0}^{\theta})$ is the radiative flux,
 ${\mathbf{P}}_{0}$ is the radiation pressure tensor,
 ${\Gamma}_{\mathrm{Comp}}$ is the energy transport rate
 from the gas to the radiation field via the Comptonization,
 $\kappa$ is the absorption opacity,
 and ${\Phi}_{\rm vis}$ is the viscous dissipation function.
We write ${\Gamma}_{\mathrm{Comp}}$ as 
\begin{equation}
{\Gamma}_{\rm Comp} =
   4{\sigma}_{\rm T}c\frac{k_{\rm B}(T_{\rm gas}-T_{\rm rad})}{m_{\rm e}c^{2}}
        \left(\frac{\rho}{m_{p}}\right)E_{0},
\end{equation}
where $T_{\rm gas}$ is the gas temperature,
 $T_{\rm rad}~[=(E_{0}/a)^{1/4}]$ is the radiation temperature,
 $a$ is the radiation constant,
 $\rho$ is the mass density of the gas,
 ${\sigma}_{\rm T}$ is the cross section for electron scattering,
 $m_{\rm e}$ is the electron mass,
 and $m_{\rm p}$ is the proton mass.
We assume a one-temperature plasma in which the electron temperature equals
 the ion temperature.
This equation is obtained by integrating the Kompaneets
 equation over the frequency (see e.g., Padmanabhan 2000 for detail).

Our methods are the same as \citet {Ohs06} except that
 we solve the energy equations including the Compton heating/cooling terms.
We employ the operator-splitting method in the energy equations
 and additionally solve the following equations;
\begin{eqnarray}
\frac{{\partial}e}{{\partial}t} &=& -{\Gamma}_{\rm Comp}~, \\
\frac{{\partial}E_{0}}{{\partial}t} &=& {\Gamma}_{\rm Comp}~,
\end{eqnarray} 
using Newton-Raphson iteration, with bisection when Newton-Raphson method fails.

We start the calculations with a hot, rarefied, optically thin atmosphere
(i.e., without an initial optically thick disk).
We solve the RHD equations numerically by an explicit-implicit 
finite-difference scheme on an Eulerian grid.
The computational domain size is
 $3r_{\rm s}~{\leq}~r~{\leq}~500r_{\rm s}$ and $0~{\leq}~{\theta}~{\leq}~{\pi}/2$.
We imposed an absorbing boundary condition at $r=r_{\rm in}=3r_{\rm s}$
by attaching a damping layer, in which the physical 
quantities $q$ gradually approach the inital values $q_0$ as
$q_{\rm new}(\rm i-n)=q({\rm i})-(q({\rm i})-q_0)f({\rm n})$,
 where $r({\rm i})=r_{\rm in}$. The function $f({\rm n})$ is a smooth function
which monotonically increases from 0 to 1 as n increases.
We continuously add mass through the outer boundary $(r=r_{\rm
out}=500r_{\rm s})$
near the equatorial plane 
$(0.45{\pi}~{\leq}~{\theta}~{\leq}~0.5{\pi})$
at a constant rate ${\dot m}_{\rm input}$. 
Here, ${\dot m}_{\rm input}$ is the mass input rate normalized by the
 critical accretion rate, ${\dot M}_{\rm crit}{\equiv}L_{\rm E}/c^{2}$.
The injected matter is assumed to have a specific angular momentum
 corresponding to the Keplerian angular momentum at $r = 100r_{\rm s}$.
On the other hand,
we allow matter to escape freely but not to enter the computational domain
at $r=r_{\rm out}$ and $0~{\leq}~{\theta}<0.45{\pi}$.
The number of grid points is $(N_{r},~N_{\theta})=(96,~96)$.
The grid points in the radial direction 
 are distributed such that ${\Delta}{\ln}~r=\textup{constant}$,
 while the grids in ${\theta}$ direction are distributed
 in such a way 
that ${\Delta}{\cos}{\theta}=1/N_{\theta}$.

\section{Numerical Results} \label{RES}
\subsection{Quasi-Steady Structure} \label{QSS}

We first overview the accretion flow and outflow properties.
The overall evolution can be classified into two phases:
a transient accumulation phase and a subsequent quasi-steady phase.
In the former one, matter injected through the outer disk boundary creates continuous 
gas inflow 
and accumulates around $r = 100r_{\rm s}$, where the angular momentum of
 the injected gas equals that of the Keplerian rotation.
Viscous processes allow the angular momentum of the gas to be 
 transported outward, which drives the gas inflow.
Eventually, the gas falls onto the black hole in a quasi-steady fashion.
In this subsection, we fix the mass input rate at the outer boundary to
a value ${\dot m}_{\rm input}=10^{3}$.

\begin{figure*}[!h]
\begin{center}
\FigureFile(150mm,100mm){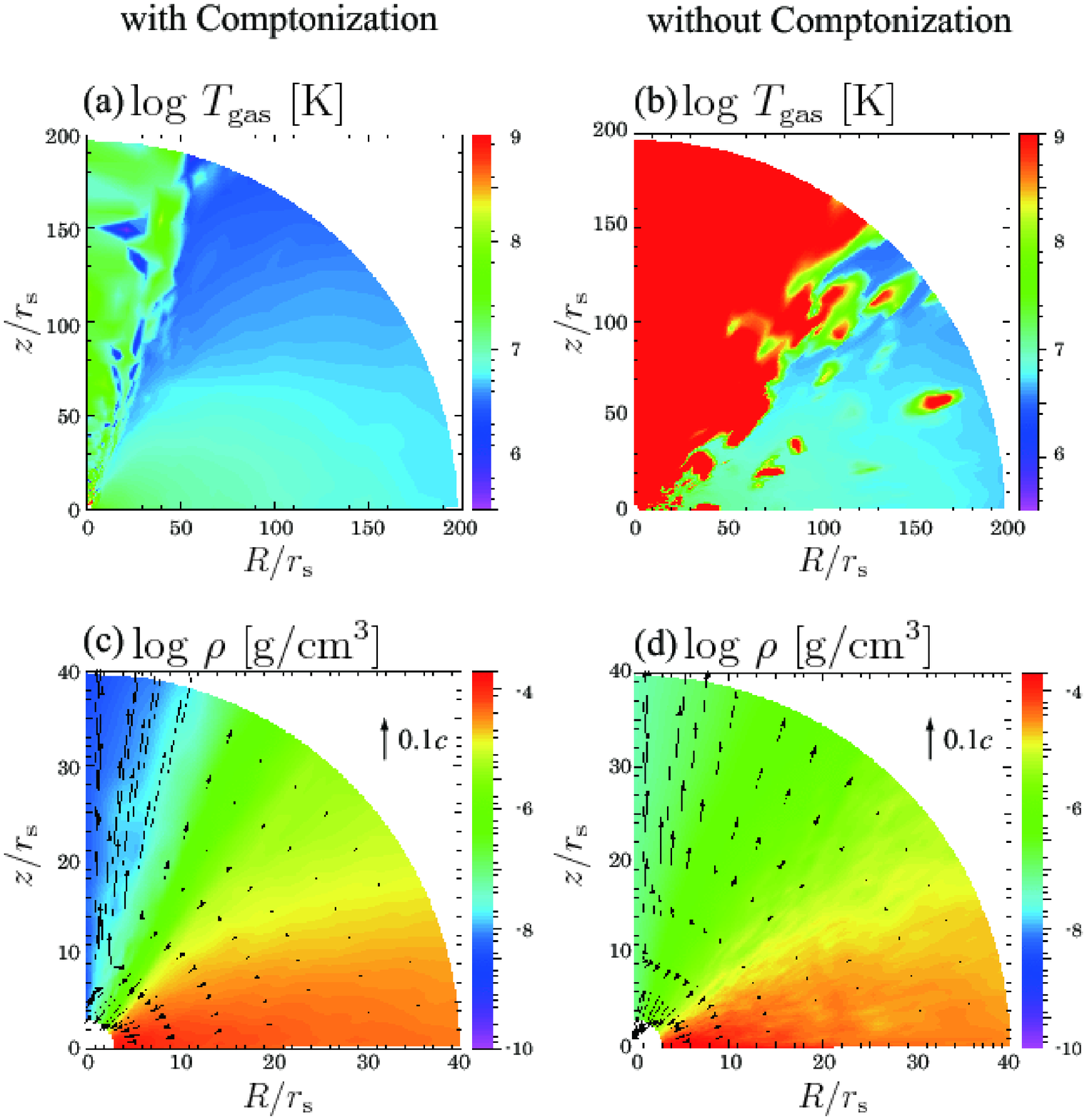} 
\caption{Distribution of temperature at $t=40\,\textup{s}$ (top) and mass density
 averaged during $t=30$--$50\,\textup{s}$ (bottom). (Left) The model including Comptonization. (Right) The model without Comptonization.
Arrows represent velocities of fluid motions.}
\label{disk} 
\end{center}
\end{figure*}

Figure \ref {disk} shows the color contours of
the gas temperature (top panels) and of the mass density overlaid with the
fluid velocity vectors (bottom panels) in the quasi-steady state for 
models with and without Comptonization in the left and right panels, respectively.
We easily notice that the gas temperature of the outflow, 
which was originally $10^{9}$--$10^{11}\,\textup{K}$ (Fig. \ref {disk}b), 
is now reduced considerably to $\sim 10^{7.5}$--$10^8\,\textup{K}$,
when we take into account the Comptonization effects (Fig. \ref {disk}a).
The gas temperature in the disk region, on the other hand, does not
change appreciably.
This is because the radiation temperature and the gas temperature are
nearly equal due to frequent absorption and emission of photons in the disk.
The ratio of the Compton cooling timescale $(t_{\rm Comp}{\sim}e/|{\Gamma}_{\rm
Comp}|)$ to the
cooling timescale by the free-free and bound-free emission
$(t_{\rm ff,bf}{\sim}e/|4{\pi}{\kappa}B-c{\kappa}E_{\rm 0}|)$
is $t_{\rm Comp}/{t_{\rm ff,bf}}{\sim}10^{-5}$ in the
outflow region and ${\sim}1$ in the disk region.
We also note that in the outflow region the ratio of the Compton cooling
timescale to the escaping timescale of the outflow ($t_{\rm esc}{\sim}r_{\rm
out}/0.1c$) is $t_{\rm Comp}/t_{\rm esc}{\sim}10^{-4}$ although that of the cooling timescale by
free-free and bound-free emission to the escaping timescale is
$t_{\rm ff,bf}/t_{\rm esc}{\sim}10^1$.
Therefore the outflowing gas is efficiently cooled by the
Comptonization before the gas escape from computational domain
 (by contraries, the cooling by free-free and bound-free emission is not
effective in the outflow region).


The outflowing gas is less dense in models including the Comptonization.
A reduction in the outflow rate leads to an increase
of the mass accretion rate onto the black hole and leads to a
geometrically thicker accretion flow.

We find that the accretion rate onto the black hole
and the total viscous heating rate in the model with Compton cooling
are larger than those in the model without Compton cooling by about $70{\%}$ and $20{\%}$, respectively.
The disk scale height is decreased by the Compton cooling in the
outflow region.  Therefore,
the gas above the disk accumulates towards the disk plane
(The radiative force in the disk region is reduced due to photon
diffusion processes).
Compared with the case without Comptonization,
the amount of the radiation-pressure driven outflow is smaller,
the mass accretion rate is larger, and the viscous heating rate is also larger
in the case with Comptonization.
The photon luminosity remains about the same, however. 
The energy of the photons which are swallowed by the black hole per unit time 
in the model with Comptonization is three times as large as that without Comptonization.
The increase in the mass accretion rate makes photon-trapping effects 
more significant \citep {Ohs02}.
Most of the photons which are produced additionally by the inclusion of
the Compton effects are generated around the innermost region of the
accretion disk (i.e., in the photon-trapping region).
Thus, they are swallowed by the black hole without escaping to the
outflow region \citep {OM07}.
This is the reason why the observed photon luminosity does not change appreciably 
by the inclusion of the Compton effects. 

\subsection{Hardness of the Photon Spectrum} \label{CYP}
The Compton cooling of the outflow region affects the radiation spectra.
In this subsection, we use time-averaged simulation results 
to discuss the spectral properties in terms of the Compton $y$-parameter, 
which indicates how much seed soft photons 
are up-scattered by the hot electrons in the outflow. 
The $y$-parameter is given by
\begin{equation}
y = \frac{4k_{\rm B}T_{\rm e}}{m_{\rm e}c^{2}}
~\max({\tau}_{\rm es},~{\tau}_{\rm es}^{2})~, \label {yp}
\end{equation}
where ${\tau}_{\rm es}$ is the Thomson optical depth of the outflow.
Note that we assume a one-temperature plasma and, hence, electron and proton temperatures are the same in this paper.

\begin{figure}
\begin{center}
\FigureFile(80mm,60mm){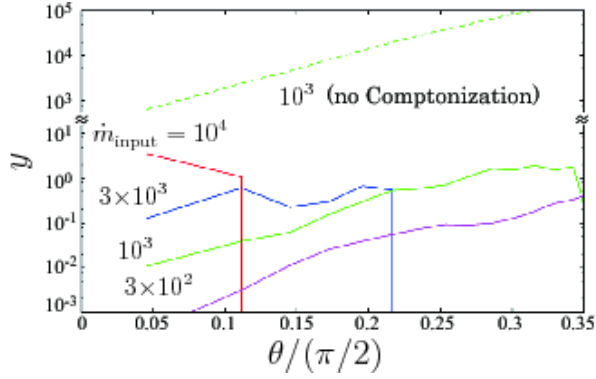}
\caption{Compton $y$-parameter in the outflow region with respect to the
 polar angle ${\theta}/({\pi}/2)$ for $\dot{m}_{\rm input}=3{\times}10^{2}$ (red), $10^{3}$ (green),
 $3{\times}10^{3}$ (blue), and $10^{4}$ (magenta).
Solid curves represent results for models with Comptonization
, the dashed curve shows the results for the model without Comptonization.
The $y$-parameters become zero when ${\theta}/({\pi}/2)~{\gtrsim}~0.12$
 and ${\gtrsim}~0.22$
 for $\dot{m}_{\rm input}=3{\times}10^{3}$ and $10^{4}$, respectively.
This is  because the gas temperature of the outflow is below $10^{7}\,\textup{K}$.
}
\label{y} 
\end{center}
\end{figure}

We calculate the Compton $y$-parameter in the following way.
First, we specify a ray, a straight line that connects a point at the outer
boundary $(r_{\rm out},{\theta})$ and the origin $(r=0)$ with
an inclination angle of $\theta$ (measured with respect to the $z$-axis).
Second, we define a photosphere for that ray by the position where 
${\tau}_{\rm eff}({\theta})={\sqrt{{\tau}_{\rm a}({\theta})({\tau}_{\rm a}({\theta})+{\tau}_{\rm es}({\theta}))}}=1$.
Here the optical depth for absorption ${\tau}_{\rm a}({\theta})$ and for
electron scattering ${\tau}_{\rm es}({\theta})$ are integrated from the
outer boundary of the computational domain.
This segment between the photosphere and the outer boundary is used to
compute the $y$-parameter.
If the medium along the ray is optically thin, we integrate the opacity
inwards to the inner boundary.
Third, we obtain the opacity-weighted mean plasma temperature along the
ray outside the photosphere by
\begin{eqnarray}
\displaystyle T_{\rm e}=\frac{\sum T_{\rm gas}(r_{\rm
 i},{\theta}){\Delta}{\tau}_{\rm es}(r_{\rm i},{\theta})}
{{\tau}_{\rm es}},
\end{eqnarray}
where ${\Delta}{\tau}_{\rm es}(r_{\rm i},{\theta})$
 is the electron scattering opacity in the i-th grid point
along the ray, and ${\tau}_{\rm es}=\sum{\Delta}{\tau}_{\rm es}(r_{\rm i},{\theta})$.
By separating the summation along the ray to the high temperature region
 up-scattering photons ($T_{\rm e}>10^{7}~{\rm K})$ and the low
 temperature region down-scattering photons ($T_{\rm e}<10^{7}~{\rm K}$),
 we confirm that the contribution to the $y$-parameter from the low
 temperature region is less than 1.
Thus we include only the grid points where $T_{\rm gas}(r_{\rm i},
 {\theta})>10^{7}~{\rm K}$ to compute $T_{\rm e}$, ${\tau}_{\rm es}$, and
 $y$ in equation \ref{yp}.
Otherwise, $y$ will be overestimated.
Figure \ref {y} shows the $y$-parameters in the outflow region.
The solid curves show the results for models with Comptonization for
various mass input rates,
whereas the dashed curve shows the $y$-parameter for a model without Comptonization.
When we include Comptonization effects in the RHD simulation,
 the $y$-parameter decreases from ${\gtrsim}10^{3}$ to $10^{-2}$--$10^{0}$
 because the gas temperature and the gas density in the outflow region
 decrease (see Fig.~\ref {disk}).
Importantly, we confirm that the Compton
$y$-parameter in models with Comptonization is consistent
 with the observations of ULXs, 
while that in models without Comptonization is too large to explain them. 
We also find that the $y$-parameter increases with an
increase of $\dot{m}_{\rm input}$ (and, hence, an increase of the photon
luminosity).
This is because the larger $\dot{m}_{\rm input}$ is,
the larger becomes the electron number density,
and, hence, the larger becomes the Thomson optical depth.
The SED becomes harder as the photon luminosity increases.

Table \ref {table} summarizes the photon index $\Gamma$, the $y$-parameter,
 the Thomson optical depth ${\tau}_{\rm es}$, the averaged
 gas temperature $T_{\rm e}$,
and the normalized isotropic (apparent) photon luminosity
 $L_{\rm ph(iso)}/L_{\rm E}$,
 for a single ray with an angle of ${\theta/({\pi}/2)}{\sim}0.05$ 
(i.e., nearly face-on view).
Here we calculate the isotropic photon luminosity by $L_{\rm ph(iso)}{\equiv}4{\pi}\left(r_{\rm
 out}\right)^{2}F^{\rm r}_{0}\left(r_{\rm out},{\theta}\right)$.
The photon index ${\Gamma}~(=-1/2+{\sqrt{9/4+4/y}})$ resulting
 from the unsaturated Comptonization \citep {RL79} is computed 
 using the $y$-parameter at ${\theta/({\pi}/2)}{\sim}0.05$.
In Table \ref {table}, we also present the normalized kinetic luminosity
 $L_{\rm kin}/L_{\rm E}$, the normalized photon luminosity
 $L_{\rm ph}/L_{\rm E}$, the mass accretion rate onto the black hole
 ${\dot m}({\equiv}{\dot M}/{\dot M}_{\rm crit})$, the mass outflow rate
 ${\dot m}_{\rm outflow}({\equiv}{\dot M}_{\rm outflow}/{\dot M}_{\rm
 crit})$, which are integrated over the all angle.
The kinetic luminosity is defined as the mechanical energy of outflow
which exceed the escape velocity $v_{\rm esc}$ at the outer boundary
$(r=r_{\rm out})$:
\begin{equation} L_{\rm kin}{\equiv} (r_{\rm out})^{2}\int \frac{1}{2}{\rho(r_{\rm out},
{\theta})}v_{r}(r_{\rm out}, {\theta})^{3}
 {\rm d}
{\Omega}
.
\end{equation}
The mass accretion rate is defined at the inner boundary as 
\begin{equation}
{\dot M}{\equiv}(r_{\rm in})^{2}\int {\rho(r_{\rm in}, {\theta})}\max[-v_{r}(r_{\rm in}, {\theta}),0]{\rm d}
{\Omega}
.
\end{equation}
The mass outflow rate is evaluated at the outer boundary:
\begin{equation}
{\dot M}_{\rm
outflow}{\equiv}(r_{\rm out})^{2}\int {\rho(r_{\rm out}, {\theta})}\max[v_{r}(r_{\rm out}, {\theta}),0]{\rm d}
{\Omega}
.
\end{equation}

\begin{figure}
\begin{center}
\FigureFile(70mm,80mm){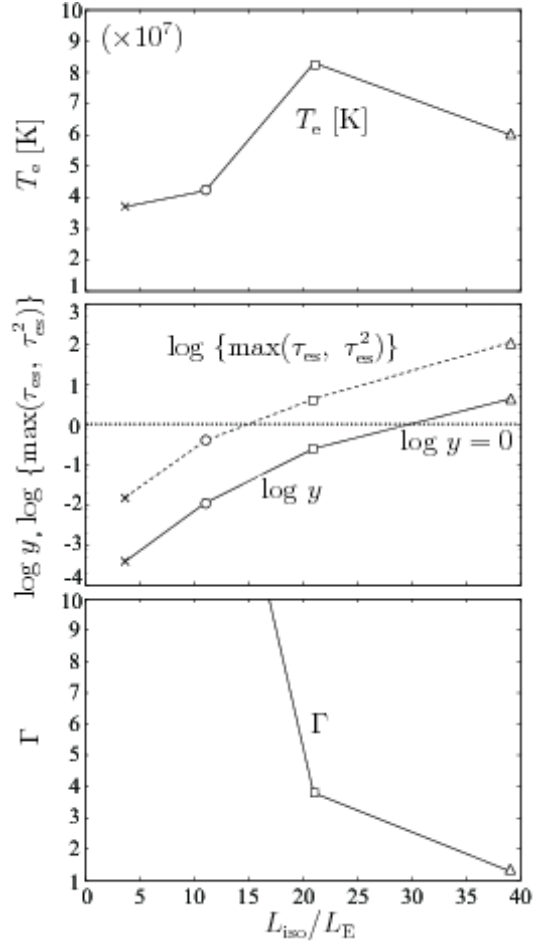}
\caption{The gas temperature of outflow averaged along the ray outside
 the photosphere $T_{\rm e}$ (top panel),
 y-parameter, Thomson optical depth ${\max}({\tau}_{\rm
 es},{\tau}_{\rm es}^2)$
 (middle panel), and the photon index ${\Gamma}$ (bottom panel) as
 functions of photon luminosity $L_{\rm ph}/L_{\rm E}$ at
 ${\theta}/({\pi}/2){\sim}0.05$. (Cross: ${\dot m}_{\rm input}=3{\times}10^{2}$, Circle: ${\dot m}_{\rm input}=10^{3}$, Square: ${\dot m}_{\rm input}=3{\times}10^{3}$, Triangle: ${\dot m}_{\rm input}=10^{4}$).
The dotted line in the middle panel represents $y=1$.}
\label{relation} 
\end{center}
\end{figure}

Figure~\ref {relation} shows the dependence of the space-averaged gas temperature
 of the outflow, the Thomson optical depth,
 the $y$-parameter, and the photon index for an observer located in the 
direction of ${\theta/({\pi}/2)}{\sim}0.05$ on the isotropic photon luminosity.
We find a negative correlation between $\Gamma$ and $L_{\rm ph(iso)}$.
This is because ${\max}({\tau}_{\rm es},~{\tau}_{\rm es}^{2})$ increases 
with an increase of ${\dot m}$.


\section{Discussion} \label{DIS}
As mentioned in $\S$ \ref{Int},
we expect a hard spectral state exists in the supercritical regimes
for the supercritical accretion model for ULXs (i.e., central black
holes are stellar-mass black holes) to be viable.
In this paper, we demonstrated that such a hard state can really
appear in the case of the supercritical accretion onto black holes
as a result of the combination
of an optically thick, slim-disk type inflow and a Comptonizing outflow.
In this new supercritical state,
higher photon luminosities correspond to harder SEDs, in analogy to black hole
binaries in the very high state, but at even higher photon luminosities.

\begin{figure*}
\begin{center}
\FigureFile(160mm,200mm){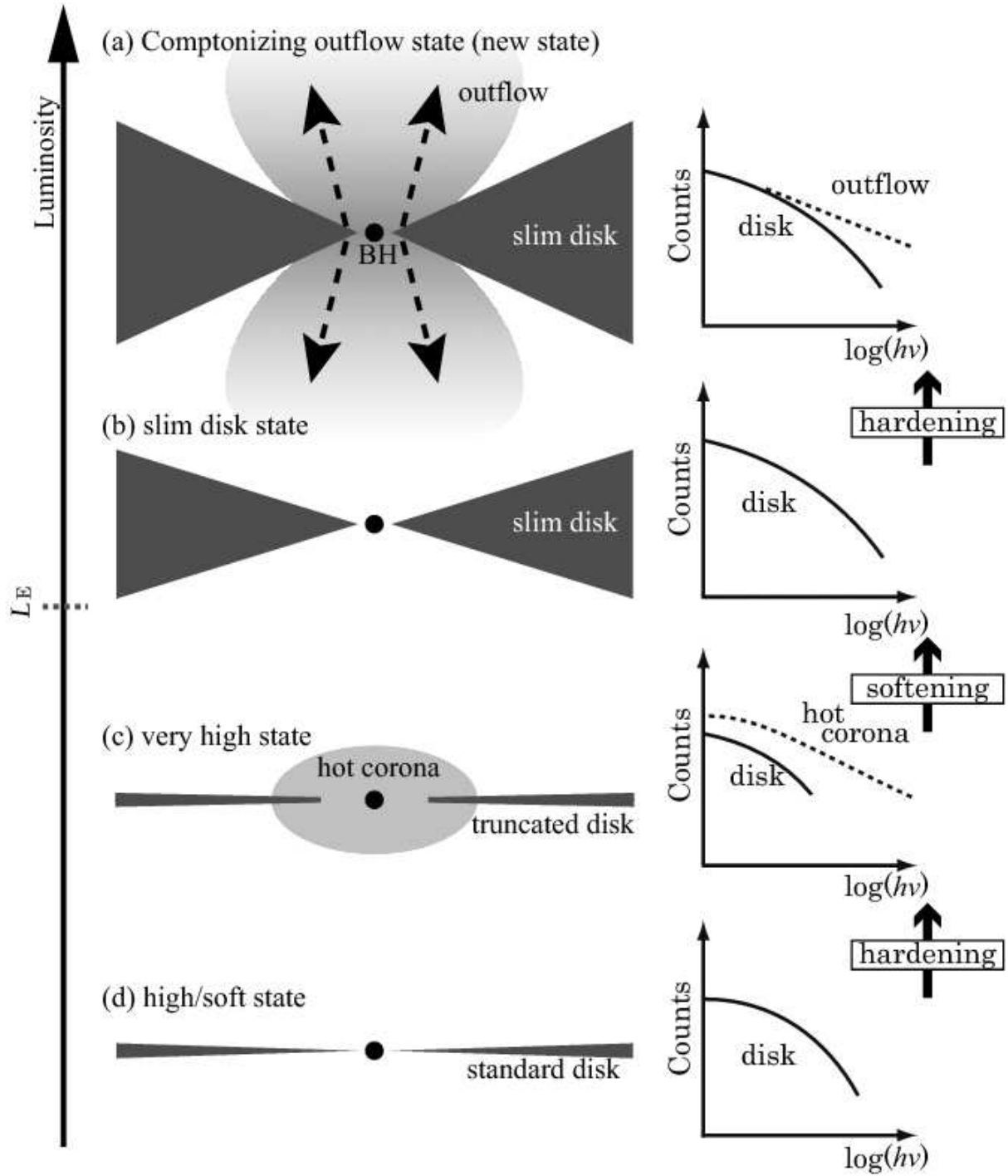}
\caption{Schematic pictures of the states of accretion disks. 
From top to bottom: (a) Comptonizing outflow state, which consists of
 supercritical accretion flows (i.e., slim disks) and Comptonizing outflows, 
(b) slim disk state, (c) very high state, and (d) high/soft state,
 The SED of the new state is harder than that of the slim disk state, because the new state includes a hot outflow which up-scatters seed
 photons from the underlying disk.}
\label{Disk_S} 
\end{center}
\end{figure*}

Figure \ref {Disk_S} illustrates schematically our proposed scenario
for the variations in the disk geometry
and the spectral properties according to the changes in the mass input rate. 
From top to bottom:
 (a) Comptonizing outflow state, (b) slim disk state,
 (c) very high state, and (d) high/soft state.
In the slim disk state, in which the photon luminosity barely exceeds $L_{\rm E}$,
the amount of gas outflow is negligible.
Hence, a pure thermal emission component is detectable.
However, as the mass input rate increases,
the amount of radiation-pressure driven, mildly hot 
(${\sim}~10^{7.5-8} {\rm K}$) outflow increases, 
thereby more soft photons from the accretion flow being Compton up-scattered.
The higher the accretion rate is, the harder becomes the SED.
This transition can account for the spectral transition reported for some
 ULXs (e.g., NGC1313 X-2), which show a positive correlation between the
 SED hardness and the photon luminosity.

What will be an observational test of our proposed spectral state?
Unfortunately, line absorption by the outflow material is
not expected, since the temperature of the Comptonizing outflow is so high
that all the metals in the Comptonizing outflow are completely ionized.
Instead, we might search for sources which show both
slim-disk features and a Comptonized blackbody component.
The slim-disk features can be evaluated through the spectral fitting
by using the extended disk-blackbody (DBB) model \citep {Mine94},
in which the temperature profile is assumed to be proportional to
$r^{-p}$, with $r$ being the distance to the central black hole
and $p$ being a fitting parameter.  The $p$-value depends on the
disk model: $p$=0.75 in the standard-type disk and $p$=0.5 in the slim disk
(see Chap. 10 of \cite {KFM08}).
Using this technique, \citet {Vi06} found that
 at least some ULXs are powered by supercritical accretion onto
 stellar-mass black holes.
If there are ULXs which show both a soft component with $p=0.5$
and a Comptonized hard component, they will provide strong support for
our model.

Next, let us discuss the effect of the bulk (motion) Comptonization.
The bulk Comptonization becomes significant when
 the speed of the bulk motion becomes relativistic.
The bulk Comptonization dominates the thermal Comptonization
when $\xi\equiv v/(12k_{\rm B}T_{\rm gas}/m_{\rm e})^{1/2}>1$ \citep {BP81}.
In our simulation results, the outflow velocity is ${\sim}~0.5~c$,
while the gas temperature is ${\sim}~10^{7.5-8}~{\rm K}$,
giving rise to $\xi \sim 1$ in the region near the polar axis.
This means that the bulk motion Comptonization is not negligible,
compared with the thermal Comptonization.
We calculate the $y$-parameter of bulk Comptonization by
 $y_{\rm bulk}=(m_{\rm e}v^{2}/3m_{\rm e}c^{2})
{\max}({\tau}_{\rm es},~{\tau}_{\rm es}^{2})$, finding that
 bulk Comptonization becomes comparable to thermal Comptonization
 in the region around the polar axis.
Hence, we expect substantial bulk Comptonization effects in the spectra
of nearly face-on sources.
Caution should be taken here, however.
Our simulations are likely to overestimate the bulk velocity of the outflow,
since the effect of the radiation drag,
 which decelerates the hydrodynamic motion, is not considered in our code.
On the other hand, we evaluate the radiation temperature
 as $T_{\rm rad}=(E_{0}/a)^{1/4}$, which results in
underestimation of the radiation temperature, 
when the radiation spectra become harder than a Planck distribution.  
Therefore,
we may overestimate the cooling rate of the gas via Comptonization,
and, hence, underestimate the gas temperature.
To evaluate the effects of bulk Comptonization more accurately,
full non-gray RHD simulations are required.
Future advances in computer technology will 
enable the study of full non-gray RHD simulations of supercritical
accretion flows.
We should also note that exponential 'photon breeding' mechanism, which converts
a kinetic energy of a relativistic jet into a radiation energy via the bulk
Comptonization by electron-positron pair plasmas in a jet (this is a
positive feedback mechanism because
pair plasma is created by high energy photons), is not efficient.
This is because the outflow speed obtained in this study is about $0.5c$
although exponential photon breeding requires highly relativistic motion
whose Lorentz factor ${\Gamma}$ exceeds 3-4 \citep {SP06}. 

 

We showed that in supercritical accretion flows
, the kinetic luminosity of the outflow 
$L_{\rm kin}$ is
$0.075,~0.30,~0.55,~1.1~L_{\rm E}$ for ${\dot m}_{\rm
input}=3{\times}10^{2},~10^{3},~3{\times}10^{3},~10^{4}$ respectively 
(see Table \ref {table}) and that the
momentum flux of the outflow is ${\sim}~2L_{\rm kin}/v~{\sim}~0.2-2~L_{\rm
E}/v~{\sim}~L_{\rm E}/c$, where $v~(\sim 0.5c)$ is the outflow velocity.
The mass, momentum, and energy outflow from the accreting
black holes can affect the surrounding interstellar gas.
This feedback has been discussed in supermassive black 
holes (AGN feedback) and should also be relevant in Galactic
microquasars. By assuming that the momentum flux of the
outflow is $\sim~L_{\rm E}/c$ as suggested by \citet
{KP03}, King (2003, 2005) showed that outflows 
from a supercritically accreting black hole at the center of the galaxy 
create a bubble, inside which stars are formed.
This mechanism can explain the relation between the black hole 
mass and the velocity dispersion of the bulge ($M-\sigma$ relation). 
The results of our simulations for ${\dot m}_{\rm input}=10^{3} ~\rm and
~3{\times}10^3$ are consistent with the 
assumptions of King (2003, 2005), if the black hole at the
galactic center is accreting the gas supercritically.
We note that compared to \citet {Ohs07b}, which performed
two-dimensional RHD simulations without Comptonization, higher mass
input rate is required in our simulations to launch the outflow with the
momentum flux $\sim L_{\rm E}/c$.
This is because the mass outflow rate decreases when Comptonization is
included, as we mentioned in \S \ref{QSS}.

In this paper we simply assumed the $\alpha$-prescription of the viscosity.
However, it is now widely accepted that
the MHD turbulence driven by the magneto-rotational instability (MRI)
provides a promising mechanism for the angular momentum transport in high-temperature
flows \citep {BH91,HGB95,BH98}.
In addition, the Parker instability, driven by the magnetic buoyancy
 \citep {Par66},
creates magnetic loops in the disk corona \citep{Mach00}.
In order to investigate supercritical accretion phenomena without
introducing the $\alpha$-viscosity, we need to perform global radiation
magnetohydrodynamic (RMHD) simulations (such
simulations are pioneered by \citet {Oh09}, but Compton effects
are not included).
To treat the relativistic effects of RHD such as
 radiation drag, it is also 
necessary to extend our numerical model of the radiative transfer to be valid up to $v/c\sim1$. This remains as future work.

\bigskip
We would like to thank T. Hanawa, S. Hirose, T. N. Kato, H. Oda,
T. Sano, H. Takahashi, and S. Takeuchi for helpful conversations.
Numerical simulations were carried out on Cray XT4 at Center for
Computational Astrophysics of National Astronomical Observatory of Japan
 and on Express5800/120Rg-1 at Cybermedia Center Osaka University.
This work is supported in part 
by Ministry of Education, Culture, Sports, Science, and 
Technology (MEXT) Young Scientist (B) 20740115 (KO), 
by the Grant-in-Aid of MEXT (19340044, SM, 20340040, RM), 
and by the Grant-in-Aid for the global COE programs on 
"The Next Generation of Physics, Spun from Diversity and Emergence" 
from MEXT (SM, DH).






\newpage

\newpage

\begin{table*} [h!]
\begin{center}
\caption{Mass accretion rate, mass outflow rate, luminosities and
 properties of Comptonizing outflows.* \label{table}}
\begin{tabular}{c|ccccccccc}
\hline\hline
${\dot m}_{\rm input}$ & ${\dot m}$ & ${\dot m}_{\rm outflow}$ &
 $L_{\rm kin}/L_{\rm E}$ & $L_{\rm ph}/L_{\rm E}$ & $L_{\rm ph(iso)}/L_{\rm E}$
 & ${\max}({\tau}_{\rm
 es},{\tau}_{\rm es}^2)$ & $T_{\rm e}$
 & $y$ & ${\Gamma}$ \\
\hline
$3{\times}10^{2}$ &$1.3{\times}10^{2}$ &$3.7{\times}10^{1}$
	 &0.075  &2.4 &3.6
 &$1.5{\times}10^{-2}$ &$3.7{\times}10^{7}$ &$3.8{\times}10^{-4}$ &100\\
$10^{3}$ &$2.4{\times}10^{2}$ &$7.5{\times}10^{2}$ &0.30
	 &3.2 &11
 &$3.9{\times}10^{-1}$ &$4.2{\times}10^{7}$ &$1.1{\times}10^{-2}$ &19\\
$3{\times}10^{3}$ &$4.9{\times}10^{2}$ &$2.5{\times}10^{3}$
	 &0.55 &4.5 &21
 &$4.4{\times}10^{0}$ &$8.3{\times}10^{7}$ &$2.5{\times}10^{-1}$ &3.8\\
$10^{4}$ &$1.3{\times}10^{3}$ &$8.6{\times}10^{3}$ & 1.1 &8.8 &39
 &$1.1{\times}10^{2}$ &$6.0{\times}10^{7}$ &$4.4{\times}10^{0}$ &1.3\\
\hline
\end{tabular}
\end{center}
* ${\dot m}_{\rm input}$ is mass input rate, $\dot m$ is mass accretion
 rate, ${\dot m}_{\rm outflow}$ is mass outflow rate, $L_{\rm kin}$ is
 kinetic luminosity, $L_{\rm ph}$ is photon luminosity, $L_{\rm
 ph(iso)}$ is isotropic photon luminosity, $L_{\rm E}$ is Eddington
 luminosity, ${\tau}_{\rm es}$ is Thomson optical depth,
 $T_{\rm e}$ is electron temperature, $y$ is Compton $y$-parameter, and
 $\Gamma$ is photon index, respectively. 
We note that ${\dot m}$, ${\dot m}_{\rm outflow}$, $L_{\rm kin}$,
and $L_{\rm ph}$ are integrated over the all angle.
By contrast, $L_{\rm ph(iso)}$ is evaluated by $L_{\rm ph(iso)}=4{\pi}(r_{\rm
 out})^{2}{\cdot}F^{r}_{0}(r_{\rm out},{\theta})$, where
 ${\theta}/({\pi}/2) {\sim} 0.05$ 
The quantities ${\max}({\tau}_{\rm es},{\tau}_{\rm es}^2)$, $T_{\rm e}$, $y$ and
 ${\Gamma}$ are also evaluated at ${\theta}/({\pi}/2){\sim}0.05$.
\end{table*}


\begin{thebibliography}{}
\bibitem[Abramowicz et al.(1988)]{Ab88} Abramowicz, M. A., Czerny, B.,
		Lasota, J. P., \& Szuszkiewicz 1988, \apj, 332, 646
\bibitem[Balbus and Hawley(1991)]{BH91} Balbus, S. A., \& Hawley, J. F. 1991, \apj, 376, 214
\bibitem[Balbus and Hawley(1998)]{BH98} Balbus, S. A., \& Hawley, J. F. 1998, Reviews of Modern Physics, 70, 1
\bibitem[Berghea et al.(2008)]{Ber08} Berghea, C. T., Wearer, K. A.,
		Colbert, E. J. M., \& Roberts, T. P. 2008, \apj, 687, 471
\bibitem[Blandford and Payne(1981)]{BP81} Blandford, R. D., \& Payne, D. G. 1981, \mnras, 194, 1033
\bibitem[Cropper et al.(2004)]{Cro04} Cropper, M., Soria, R.,
		Mushotzky, R. F., Wu, K., Markwardt, C. B., \& Pakull,
		M. 2004, \mnras, 349, 39
\bibitem[Dolan et al.(1979)]{Dol79} Dolan, J. F., Grannell, C. J.,
		Dennis, B. R., Frost, K. J., \& Orwig, L. E. 1979, \apj,
		230, 551
\bibitem[Ebisawa et al.(2003)]{Ebi03} Ebisawa, K., ${\dot {\rm Z}}$ycki,
		P., Kubota, A., Mizuno, T.,  \& Watarai, K. 2003, \apj, 597, 780
\bibitem[Eggum et al.(1988)]{Egg88} Eggum, G. E., Coroniti, F. V., \& Katz, J. I. 1988, \apj, 330, 142
\bibitem[Esin et al.(1997)]{Esi97} Esin, A. A., McClintock, J. E., \&
		Narayan, R. 1997, \apj, 489, 865
\bibitem[Hawley et al.(1995)]{HGB95} Hawley, J. F., Gammie, C. F., \& Balbus, S. A. 1995, \apj, 440, 742
\bibitem[Heinzeller et al.(2006)]{HMO06} Heinzeller, D., Mineshige, S.,
		\& Ohsuga, K. 2006, \mnras, 372, 1208
\bibitem[Kato et al.(2008)]{KFM08} Kato, S., Fukue, J., \& Mineshige,
		S. 2008, Black-Hole Accretion Disks Towards a New
		Paradigm (Kyoto: Kyoto University Press)
\bibitem[Kajava \& Poutanen.(2008)]{KP08} Kajava, J. J. E., \&
		Poutanen, J.:astro-ph/0809.4634(2008)
\bibitem[Kubota and Makishima(2001)]{KM01} Kubota, A., \& Makishima, K. 2001,
		\apj, 547, 119
\bibitem[King et al.(2001)]{King01} King, A. R., Davies, M. B., Ward,
		M. J., Fabbiano, G., \& Elvis, M. 2001,
		\apj, 552, 109
\bibitem[King(2003)]{King03} King, A. R. 2003,
		\apj, 596, 27
\bibitem[King and Pounds(2003)]{KP03} King, A. R., \& Pounds, K.A. 2003,
		\mnras, 345, 657
\bibitem[King(2005)]{King05} King, A. R. 2005,
		\apj, 635, 121
\bibitem[Kubota and Done(2004)]{KD04} Kubota, A., \& Done, C. 2004,
		\mnras, 353, 980
\bibitem[Levermore \& Pomraning(1981)]{LP81} Levermore, C. D., \& Pomraning, G. C. 1981, \apj, 248, 321
\bibitem[Makishima et al.(2000)]{Mak00} Makishima, K. et al. 2000, \apj, 535, 632
\bibitem[Machida et al.(2000)]{Mach00} Machida, M., Hayashi, M., \& Matsumoto, R. 2000, \apj, 532, 67
\bibitem[Miller et al.(2003)]{Mil03} Miller, J. M., Fabbiano, G.,
		Miller, M. C. \& Fabian, A. C. 2003, \apj, 585, 37
\bibitem[Miller et al.(2004)]{Mil04} Miller, J. M., Fabian, A. C., \&
		Miller, M. C. 2004, \apj, 614, 117
\bibitem[Mineshige et al.(1994)]{Mine94} Mineshige, S., Hirano, A.,
		Kitamoto, S., Yamada, T., \& Fukue, J. 1994, \apj, 426, 308
\bibitem[Miyamoto et al.(1991)]{Miya91} Miyamato, S., Kimura, K.,
		Kitamoto, S., Dotani, T., \& Ebisawa, K. 1991, \apj, 383, 784
\bibitem[Mizuno et al.(2007)]{Miz07} Mizuno, T. et al. 2007, \pasj, 59, 257
\bibitem[Ohsuga et al.(2002)]{Ohs02} Ohsuga, K., Mineshige, S., Mori, M., \& Umemura, M. 2002, \apj, 574, 315
\bibitem[Ohsuga et al.(2005)]{Ohs05} Ohsuga, K., Mori, M., Nakamoto, T., \& Mineshige, S. 2005, \apj, 628, 368
\bibitem[Ohsuga(2006)]{Ohs06} Ohsuga, K. 2006, \apj, 640, 923
\bibitem[Ohsuga(2007a)]{Ohs07} Ohsuga, K. 2007a, \pasj, 59, 1033
\bibitem[Ohsuga(2007b)]{Ohs07b} Ohsuga, K. 2007b, \apj, 659, 205
\bibitem[Ohsuga and Mineshige(2007)]{OM07} Ohsuga, K., \& Mineshige,
		S. 2007, \apj, 670, 1283
\bibitem[Ohsuga et al.(2009)]{Oh09} Ohsuga, K., Mineshige, S., Mori, M., \&
		Kato, Y. 2009, \pasj, in press
\bibitem[Okajima et al.(2006)]{Oka06} Okajima, T., Ebisawa, K., \& Kawaguchi, T. 2006, \apj, 652, 1050
\bibitem[Okuda(2002)]{Oku02} Okuda, T. 2002, \pasj, 54, 253
\bibitem[Okuda et al.(2005)]{Oku05} Okuda, T., Teresi, V., Toscano, E.,
		\& Molteni, D. 2005, \mnras, 357, 295
\bibitem[Paczy\'nsky \& Wiita(1980)]{Pacz80} Paczy\'nsky, B., \& Wiita, P. J. 1980, \aap, 88, 23
\bibitem[Padmanabhan(2000)]{Pad00} Padmanabhan, T. 2000, Theoretical
		Astrophysics Vol 1: Astrophysical processes (Cambridge: Cambridge
		Univ. Press)
\bibitem[Parker(1966)]{Par66} Parker, E. N. 1966, \apj, 145, 811
\bibitem[Poutanen et al.(2007)]{Pou07} Poutanen, J., Lipunova, G.,
		Fabrika, S., Butkevich, A. G., \& Abolmasov, P. 2007,
		\mnras, 377, 1187
\bibitem[Roberts et al.(2006)]{Rob06} Roberts, T. P., Kilgard, R. E.,
		Warwick, R. S., Goad, M. R., \& Ward, M. J. 2006,
		\mnras, 371, 1877
\bibitem[Rybicki \& Lightman(1979)]{RL79} Rybicki, G. B., \& Lightman,
		A. P. 1979, Radiative Processes in Astrophysics (New
		York: Wiley)
\bibitem[Shakura \& Sunyaev(1973)]{SS73} Shakura, N. I., \& Sunyaev, R. A. 1973, \aap, 24, 337
\bibitem[Stern \& Poutanen(2006)]{SP06} Stern, B. E., \& Poutanen, J. 2006, \mnras, 372, 1217
\bibitem[Tananbaum et al.(1972)]{Tan72} Tananbaum, H., Gursky, H., Kellogg,
		E., Giacconi, R., \& Jones, C. 1972,
		\apj, 177, 5
\bibitem[Tsunoda et al.(2006)]{Tsu06} Tsunoda, N., Kubota, A., Namiki,
		M., Sugiho, M., Kawabata, K., \& Makishima, K. 2006,
		\pasj, 58, 1081
\bibitem[Turner \& Stone(2001)]{TS01} Turner, N. J., \& Stone, J. M. 2001, \apjs, 135, 95
\bibitem[Vierdayanti et al.(2006)]{Vi06} Vierdayanti, K., Mineshige, S., Ebisawa, K., \& Kawaguchi, T. 2006, \pasj, 58, 915
\bibitem[Vierdayanti et al.(2008)]{Vi08} Vierdayanti, K., Watarai, K., \& Mineshige, S. 2008, \pasj, 60, 653
\bibitem[Watarai et al.(2000)]{Wa00} Watarai, K., Fukue, J., Takeuchi, M., \& Mineshige, S. 2000, \pasj, 52, 133


\end{thebibliography}
\end{document}